\documentstyle[12pt,aasms4,flushrt]{article}

\newcommand{\kms}{~{\rm km\ s}^{-1}}

\newcommand{\qno}{q_0}
\newcommand{\hno}{{\rm H}_0}

\begin{document}
\title{Limits on the Gunn-Peterson Effect at \lowercase{$z$} = 5}
\author{Antoinette Songaila\altaffilmark{1}, Esther M. Hu\altaffilmark{1}, 
  Lennox L. Cowie\altaffilmark{1}}
\affil{Institute for Astronomy, University of Hawaii, 2680 Woodlawn
  Drive, Honolulu, HI 96822}
\and
\author{Richard G. McMahon}
\affil{Institute of Astronomy, University of Cambridge, Madingley Road,
  Cambridge CB3\thinspace{0HA}}
\altaffiltext{1}{Visiting Astronomer, W. M. Keck Observatory,
  jointly operated by the California Institute of Technology, the University
  of California, and the National Aeronautics and Space Administration.}

\begin{abstract}
We report new limits on the Gunn-Peterson effect at a redshift near 5 using
spectroscopic observations of the $z=5$ Sloan Digital Sky Survey quasar,
J033829.31+002156.3, made with the LRIS and HIRES spectrographs on the Keck
telescopes.  Lower resolution spectrophotometrically calibrated observations
made with LRIS over the wavelength region $\lambda\lambda4500-9600$ \AA\ were
used to obtain a continuum shape and to flux calibrate much higher resolution
($R=36,000$) observations made with HIRES.  The LRIS data show an Oke $D_A$
index of 0.75.  Portions of the HIRES spectrum return to near the extrapolated
continuum level.  Including both statistical and systematic errors we place an
upper limit of $\tau = 0.1$\ on the regions of minimum opacity.  We argue
that, even if this opacity arises in underdense regions of the intergalactic
gas, we require a high value of the metagalactic ionizing flux at these
redshifts ($J_{\nu} \gg 4 \times 10^{-23}~{\rm ergs\ cm^{-2}\ s^{-1}\ Hz^{-1}\
sr^{-1}}$ at $z\sim4.72$) to produce a solution which is consistent with even
minimum nucleosynthesis estimates of the baryon density.  We also report the
presence of a \ion{Mg}{2} absorption system of extremely high equivalent width
($W_{\lambda,rest}(2796)=1.73$ \AA) at $z=2.304$.
\end{abstract}


\section{Introduction}

Soon after the discovery of cosmologically distant objects it was realised
\markcite{field59}(Field 1959) that the absence of strong Lyman alpha
scattering of light at the appropriate redshifted wavelengths placed strong
constraints on the amount of neutral hydrogen in the intergalactic medium
\markcite{shklov64,scheuer65,gunnp}(Shklovskij 1964; Scheuer 1965; Gunn \&
Peterson 1965).  The absence of significant neutral opacity in a uniformly
distributed component of the IGM is generally referred to as the Gunn-Peterson
effect.

While it remains conceptually possible that there is a diffuse substrate of
the IGM in a multi-phase gas, modern interpretations of the Ly$\alpha$\ forest
of neutral hydrogen absorption lines suggest instead that the IGM is a highly
structured warm ionized gas in which the amplitude of the perturbations grows
gravitationally.  In this interpretation the Ly$\alpha$\ forest is produced by
scattering from the very small neutral hydrogen fraction of this undulating
density intergalactic gas \markcite{cen94,zhang95}(e.g., Cen {et~al.} 1994;
Zhang, Anninos, \& Norman 1995).  Regions of minimum \ion{H}{1} optical depth,
in this scenario, occur in cosmic mini-voids, which are underdense expanding regions
where growth is essentially frozen \markcite{meiksin94,reisen95}({Meiksin}
1994; Reisenegger \& {Miralda-Escud\'e} 1995).  The cosmic mini-voids are the
closest analogs in this model to the Gunn-Peterson effect in a homogeneous
diffuse gas, and the simple physics of these regions makes them powerful probes
of the metagalactic ionizing flux relative to the baryon density of the
universe.

Because of the rapid increase in the gas density with redshift, it is expected
that the fraction of the quasar spectrum returning to the continuum level will
diminish rapidly at the higher redshifts.  As a specific example,
\markcite{zhang97}Zhang {et~al.} (1997) find, at $z = 5$, for $\hno = 50~{\rm
km\ s^{-1}\ Mpc^{-1}}$, $\Omega_b = 0.06$\ and a Haardt-Madau \markcite{hm}
(1996) spectrum and ionizing flux, that less than a fraction of a percent of
the spectrum will have $\tau({\rm H~I}) < 0.1$.  Moreover, they find that the
spectrum of a $z = 5$\ quasar will have an average flux in the region between
$1050$~\AA\ and $1170$~\AA\ that is 10\% of the continuum value which would be
present in the absence of Lyman alpha scattering.

The newly discovered $z = 5$\ quasar J033829.31 + 002156.3
\markcite{fan99}(Fan {et~al.} 1999) presents an opportunity to look at this
question at the highest redshift yet available, and we report here low and
high resolution spectral observations of this object.  As
\markcite{ssg91b}Schneider, Schmidt, \& Gunn (1991b) found in the $z = 4.897$\
quasar PC1247+3406, $1-D_A$\ (the Oke index $D_A$\ is defined in \S 3) is
higher than the model values (0.25 in J033829.31 + 002156.3 and 0.36 in
PC1247+3406) and we also find that there are significant regions of low
optical depth ($\tau < 0.1$).  Our results are similar to the minimum
optical depth of $\tau =0.02 \pm 0.03$ at $z=4.3$ found in the $z=4.7$ quasar
BR1202--0725 by \markcite{giallo94}{Giallongo} {et~al.}  (1994) and to the $2\
\sigma$ lower limit of 0.1 found by \markcite{williger94}Williger {et~al.}
(1994) in the $z=4.5$ quasar BR1033--0327. Both the Oke index and the low
minimum opacity require a relatively high ionizing flux 
at these redshifts.

\section{Observations}

Long-slit spectroscopic observations of J033829.31 + 002156.3 were obtained
with LRIS \markcite{lris}(Oke {et~al.} 1995) on UT 10 Feb 1999 with a
$1\farcs0$ wide slit and $400\ell$/mm grating blazed at 8500 \AA\ (spectral
resolution 8.1 \AA), and on UT 17 Feb 1999 with a $1\farcs5$ wide slit, with
both the $400\ell$/mm grating blazed at 8500 \AA\ (spectral resolution 12.3
\AA), and with the $300\ell$/mm grating blazed at 5000 \AA\ (spectral
resolution 17.3 \AA). Seeing was $0\farcs7-0\farcs9$ FWHM on the first night,
and $0\farcs7-1\farcs0$ on the second night; observations were made under dark,
photometric conditions in each instance. A sequence of three exposures, shifted
by 10$''$ along the slit, was taken in each configuration with the position
angle of the slit set to the parallactic angle for the middle exposure.  This
yielded net integrations of 3600 s (5542--9350 \AA; 8.1 \AA\ resolution;
$1\farcs0$ slit), 3600 s (5799--9609 \AA; 12.4 \AA\ resolution; $1\farcs5$
slit), and 3000 s (3673--8708 \AA; 17.3 \AA\ resolution; $1\farcs5$ slit).  A
GG495 blocking filter was used to suppress second order blue light.  The
observations are summarized in Table~\ref{tbl-1}.  Calibration stars (HZ4 and
HZ44) were observed at the parallactic angle for each slit/grating
configuration and used to flux calibrate the data.  
The calibration stars used for the LRIS and HIRES data are
Category 1 HST Standard Stars \markcite{hst_std}(Bohlin \& Lindler 1992), which
are part of an ongoing effort to develop a consistent set of fundamental flux
standards for observations with the {\it Hubble Space Telescope\/} over the
wavelength region from 1050--10000 \AA\ based on the \markcite{oke90}Oke (1990)
measurements taken on the Double Spectrograph of the Palomar 5-m telescope for
optical measurements.  Although both HZ4 and HZ44 have long-wavelength flux
calibration data measured by a variety of authors
\markcite{oke74,stone77,massey90}(e.g., Oke 1974; Stone 1977; Massey \&
Gronwall 1990), for self-consistency we use the Oke \markcite{oke74,oke90}
(1974, 1990) measurements. The estimated galactic extinction is $E_{B-V}$ =
0.10 in this direction \markcite{schlegel98}(Schlegel, Finkbeiner, \& Davis
1998).

The HIRES \markcite{vogt}(Vogt {et~al.} 1994) observations were obtained using
the red collimator and the D1 decker ($1\farcs1\ \times 14''$ slit; resolution
$R=36,000$) on the nights of UT 14 and 15 Feb 1999 under conditions of variable
seeing.  The four 2400 s exposures were taken at ${\rm P.A.} = 100\arcdeg$
(non-parallactic angle) using the atmospheric dispersion corrector to maintain
the guide object for this observation within the guide camera field.  The white
dwarf standard star G191-B2B \markcite{oke90}(Oke 1990) was used to correct the
continuum fit, which was then flux-calibrated by comparing an appropriately
smoothed spectrum with the lower resolution LRIS data.  The combined LRIS
spectra, smoothed to the lowest resolution data taken with the $300\ell$/mm
grating (resolution 17.4 \AA), is shown in Figure~\ref{fig:1}, and the HIRES
spectrum in Figure~\ref{fig:2}.  Prior to any extinction corrections, both
continua are well fit by a flat $f_{\nu}$\ spectrum in the line-free portions
of the red spectrum, and this is shown by the dashed line.

The fluxed LRIS spectra are in excellent agreement with the photometric
measurements of Fan et al.\ \markcite{fan99} (1999), and with their estimated
$AB_{1450}$ continuum magnitude of 20.01 (35 $\mu$Jy).  A strong \ion{Mg}{2}
doublet is visible superposed on the \ion{C}{4} emission (Figure~\ref{fig:1}).
We measure a doublet ratio $D_R = 1.25$ and a redshift $z=2.304$ for this
system.  If the strong rest equivalent width ($W_{{\lambda},\,rest}(2796) =
1.73$ \AA) indicates a large galaxy along the line-of-sight
\markcite{Bergeron91, stei94}(Bergeron \& Boiss\'e 1991; Steidel, Dickinson, \&
Persson 1994), the quasar may be amplified by lensing.

\section{Discussion}

The simplest measures of the opacity of the neutral hydrogen are the Oke and
Korycanski \markcite{okekor} (1982) indices.  Following
\markcite{ssg91a,ssg91b}Schneider, Schmidt, \&  Gunn (1991a) and  Schneider
{et~al.} (1991b) we measure the index $$D_A = \left\langle 1 - {{f_{\nu}({\rm
observed})} \over {f_{\nu}({\rm continuum})}} \right\rangle$$ in the rest-frame
wavelength range $1050$ \AA\ to $1170$ \AA, obtaining $D_A = 0.75$\ in both the
lower resolution spectra.  This can be compared with the values of 0.55 to 0.74
measured by \markcite{ssg91b}Schneider {et~al.} (1991b) and
\markcite{kennefick95b}Kennefick, Djorgovski, \& {de  Carvalho} (1995) in
quasars in the range $z = 4.35 - 4.5$\ and with the values of 0.74 obtained by
Schneider et al.\ for the $z = 4.733$\ quasar PC1158+4635 and 0.64 for the $z =
4.897$\ quasar PC1247+3406.  It is considerably smaller than the value of 0.9
predicted by \markcite{zhang97}Zhang {et~al.} (1997) for models with the Haardt
and Madau \markcite{hm} (1996) $J_{\nu}$\ evolution, with $\Omega_b =
0.06$\ and $\hno = 50\kms\ {\rm Mpc}^{-1}$, suggesting that these models are
underestimating $J_{\nu}/\Omega_b^2$\ at high redshift.

This can also be examined by looking at the regions of minimum opacity in the
much higher resolution HIRES spectrum.  As can be seen from Figure~\ref{fig:2},
portions of the HIRES spectrum return close to the extrapolated continuum.
14\% of the spectrum has $\tau < 0.1$\ in the observed wavelength range $6800 -
7000$ \AA\ if we extrapolate the continuum with a $\nu^0$\ power law.  This
fraction is not highly sensitive to the power law choice, and reduces only to
11\% if we use a $\nu^1$\ extrapolation.  The regions of minimum opacity around
6790 \AA\ and 6948~\AA\ have extremely low optical depth.  The mean optical
depth between 6946 \AA\ and 6949 \AA\ is $-0.05 \pm 0.05$\ for a
$\nu^0$\ extrapolation and $0.05 \pm 0.05$\ for a $\nu^1$\ extrapolation, where
the errors are based only on the statistical noise in the region.  In the
subsequent discussion we will adopt what we believe is a conservative limit of
$\tau < 0.1$\ at this redshift of 4.72.

For a uniformly distributed IGM and $\qno = 0.5$,
$$\tau = 1400\,h_{50}^3\,\Omega_b^2\,J_{-22}^{-1}\,T_4^{-0.75}\,\left ( {{1+z}
\over {5.72}} \right ) ^{4.5}$$
where $h_{50}$\ is the Hubble constant in units of $50\kms\ {\rm Mpc}^{-1}$,
$J_{-22}$\ is the metagalactic flux at the Lyman edge ($J_{\nu} =
10^{-22}J_{-22}(\nu /\nu_{912})^{-\alpha}\ {\rm ergs\ cm^{-2}\ s^{-1}\
Hz^{-1}\ sr^{-1}}$\ with $\alpha = 0.7$ ) and $T_4$\ is the temperature of
the gas in units of $10^4$~K \markcite{giallo94}({Giallongo} {et~al.} 1994).  For $\tau < 0.1$\ we obtain
$$\Omega_b\,h_{50}^2 < 8.4\times
10^{-3}\,h_{50}^{0.5}\,J_{-22}^{0.5}\,T_4^{0.375}.$$
Using Haardt \& Madau's value
of $J_{-22} = 0.4$\ at $z = 4.72$\ would give $\Omega_b h_{50}^2 <
0.005\,h_{50}^{0.5}\,T_4^{0.375}$.

If, however, the minimum opacity arises in cosmic minivoids, this number must
be corrected upward to allow for the underdensity in these regions.
Simulations with current cosmological parameters suggest that the profile of
the density distribution at $z = 5$\ will be centered on a mode of an
underdensity of about 0.5, and will have very little volume indeed at
underdensities of 0.2 or less \markcite{zhang98}(e.g., Zhang {et~al.} 1998).
The underdense regions are relatively cold, with temperatures of around
5000~K.  If we conservatively assume that the minimum opacity portions of the
current spectrum arise in regions which are underdense by a factor of five,
and set $T_4 = 0.5$, we would obtain $\Omega_bh_{50}^2$\ of 0.019 for $J_{-22}
= 0.4$.  We can compare this estimate of the baryon density to that derived
from measurements of primordial $D/H$.  The lowest value of
$\Omega_b\,h_{50}^2$\ (0.02) that might currently be possible would
correspond to $D/H = 2\times 10^{-4}$ \markcite{deut_limits, webb}(Songaila,
Wampler, \& Cowie 1997; Webb et al.\ 1997) whereas $D/H = 3.3 \times 10^{-5}$, or
$\Omega_b\,h_{50}^2 = 0.077 \pm 0.006$, is obtained by Tytler's group
\markcite{burles98}(Burles \& Tytler 1998).  Even with the minimum
$\Omega_b\,h_{50}^2 = 0.02$, we require $J_{-22} >0.4$\ at $z = 4.72$, whereas
the low $D/H$\ value would require $J_{-22} >7$, more than an order of
magnitude higher than the Haardt-Madau estimate.

\acknowledgements

We thank T.\ Bida, R.\ Goodrich, G.\ Wirth, J.\ Aycock, C.\ Sorenson, and
W.\ Wack for their assistance in obtaining the observations and M. Rauch and
A. Meiksin for helpful conversations.  This work was
supported in part by the State of Hawaii and by NSF grant AST 96-17216 and NASA
grant GO-7266.01-96A from the Space Telescope Science Institute, which is
operated by AURA, Inc., under NASA contract NAS 5-26555.  R.G.M. thanks the
Royal Society for support.


\clearpage
%
%
\begin{deluxetable}{ccccccccc} 
\tablecolumns{9} 
\tablewidth{0pc} 
\tablecaption{Spectroscopic Observations of J033829.31+002156.3\label{tbl-1}}
\tablehead{ 
\colhead{}             & \colhead{}                                        &
\colhead{$t_{exp}$}    & \colhead{$\lambda\lambda$ Range\tablenotemark{a}} &
\colhead{Slit}         & \colhead{PA\tablenotemark{b}}                     & 
\colhead{Resolution\tablenotemark{c}}    & \colhead{}                      & 
\colhead{FWHM\tablenotemark{e}} \\
\colhead{UT Date}               & \colhead{A.M.}                           &
\colhead{(s)}                   & \colhead{(\AA)}                          &
\colhead{(arcsec)}              & \colhead{($^{\circ}$)}                   &
\colhead{(\AA)}                 & \colhead{Grating\tablenotemark{d}}       &
\colhead{(arcsec)}
}
\startdata 
\sidehead{LRIS}
10 Feb 1999 & 1.08 & 3600 & 5542--9350 & $1.0 \times 274$ & 28.0 & 8.1  
& 400/8500 & 0.72 \\
17 Feb 1999 & 1.12 & 3600 & 5799--9609 & $1.5 \times 274$ & 50.0 & 12.4  
& 400/8500 & 0.71 \\
17 Feb 1999 & 1.30 & 3000 & 3673--8708 & $1.5 \times 274$ & 62.0 & 17.3 
& 300/5000 & 1.00 \\
\sidehead{HIRES}
14 Feb 1999 & 1.15 & 2400 & \nodata & $1.1 \times 14$ & 100.0 & \nodata & Red 
& 3.0 \\
15 Feb 1999 & 1.10 & 7200 & \nodata & $1.1 \times 14$ & 100.0 & \nodata & Red 
& 1.2 \\
\enddata 
\tablenotetext{a}{The usable wavelength range for the multi-order HIRES
 data is $\sim6800-8000$\AA}
\tablenotetext{b}{Set close to estimated parallactic angle at mid-exposure 
 in the sequence of LRIS observations}
\tablenotetext{c}{$R=36000$ for HIRES}
\tablenotetext{d}{Identified by ruled $\ell$/mm and blaze wavelength in
 \AA\ for LRIS; for HIRES this is the RED configuration with D1 decker}
\tablenotetext{e}{Measured FWHM of quasar profile on slit}
\end{deluxetable} 

\clearpage
%
%
\begin{deluxetable}{ccc}
\tablecolumns{3}
\tablewidth{0pc}
\tablecaption{$z=2.304$ Mg~II Absorber\label{tbl-2}}
\tablehead{
\colhead{}          & \colhead{$\lambda$}     &
\colhead{$W_{\lambda,\,rest}$} \\
\colhead{ID}        & \colhead{(\AA)}         & \colhead{(\AA)}
}
\startdata
{\ion{Mg}{2}} 2796 & 9236.7 & 1.73 \\
{\ion{Mg}{2}} 2803 & 9260.4 & 1.38 \\
\enddata
\end{deluxetable}

\clearpage

\newpage

\figcaption{Both the $300\ell$/mm and $400\ell$/mm fluxed LRIS spectra are
shown, in $\mu$Jy $versus$\ wavelength.  The $300\ell$/mm spectrum is shown
over the wavelength range $4000 - 8200$ \AA\ and the combined $400\ell$/mm
data from 5700 \AA\ to 9700 \AA.  The spectra agree extremely well in the
overlap region.  The dashed and dotted lines show $\nu^0$\ and $\nu^1$\
continuum fits, normalized to the region $7700 - 8000$ \AA, which lies redward
of the atmospheric A band but blueward of the Si~IV emission line.
\label{fig:1}
}
\figcaption{The flux-calibrated HIRES spectrum over the range $6800 - 7700$ \AA.
The spectrum has been smoothed to a resolution of 2 \AA.  The dotted and
dashed lines show the same continuum fits as in Figure~1.
\label{fig:2}
}

\clearpage

\begin{figure}
\figurenum{1}
\plotone{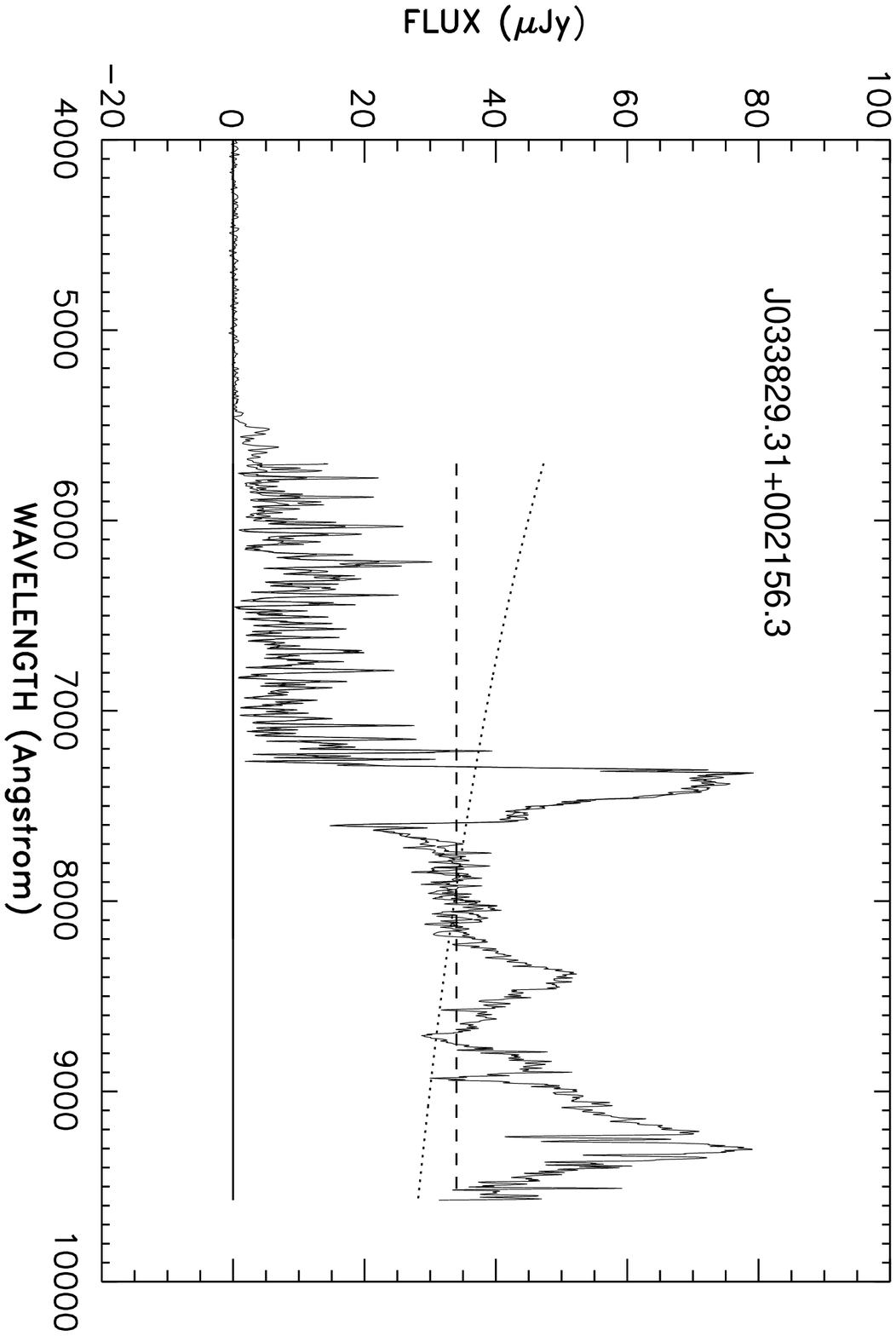}
\caption{}
\end{figure}

\begin{figure}
\figurenum{2}
\plotone{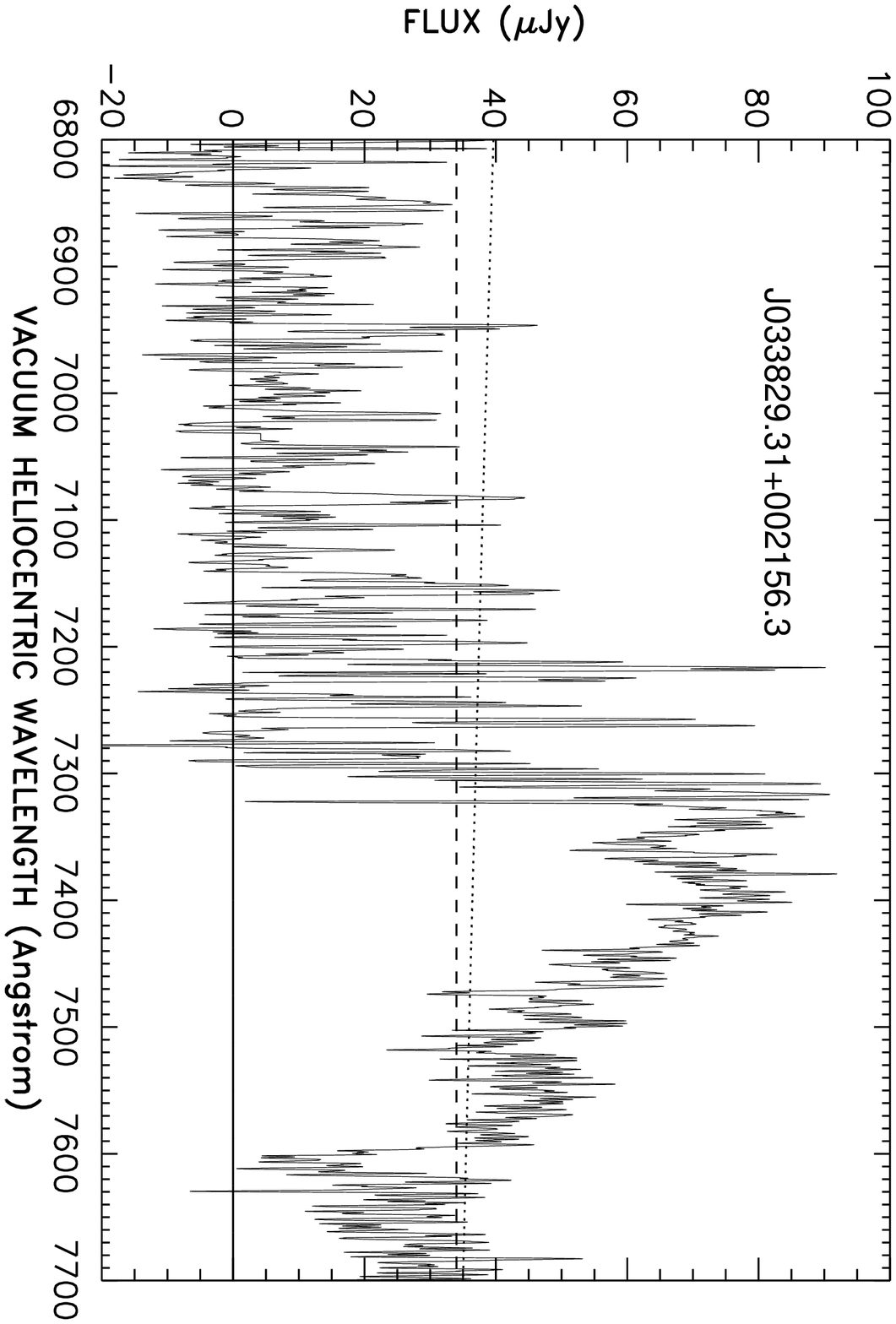}
\caption{}
\end{figure}

\clearpage


\begin{thebibliography}{}

\bibitem[Bergeron \& Boiss\'e (1991)]{Bergeron91}
Bergeron, J., \& Boiss\'e, P. (1991), \aap, 243, 344

\bibitem[Bohlin \& Lindler (1992)]{hst_std}
Bohlin, R., \& Lindler, D. 1992, Space Telescope Science Institute Newsletter, 
9, No.\ 2, 19

\bibitem[Burles \& Tytler (1998)]{burles98}
Burles, S., \& Tytler, D. 1998, \apj, 499, 699

\bibitem[Cen, {Miralda-Escud\'e}, Ostriker, \&  Rauch (1994)]{cen94}
Cen, R., {Miralda-Escud\'e}, J., Ostriker, J.~P., \& Rauch, M. 1994, \apjl,  
437, L9

\bibitem[Fan, Strauss, Schneider, Gunn, Lupton, Yanny,  Anderson, {Anderson,
Jr.}, Annis, Bahcall, Bakken, Bastian, {et~al.} (1999)]{fan99} 
Fan, X.,  et~al.  1999, \aj, 118, in press, astro-ph/9903237

\bibitem[Field (1959)]{field59}
Field, G.~B. 1959, \apj, 129, 536

\bibitem[Giallongo, {D'Odorico}, Fontana,  McMahon, Savaglio,
Cristiani, Molaro, \& Trevese (1994)]{giallo94} 
Giallongo, E., {D'Odorico}, S., Fontana, A., McMahon, R.~G., Savaglio, S.,
Cristiani, S., Molaro, P., \& Trevese, D. 1994, \apjl, 425, L1

\bibitem[Gunn \& Peterson (1965)]{gunnp}
Gunn, J.~E., \& Peterson, B.~A. 1965, \apj, 142, 1633

\bibitem[Haardt \& Madau (1996)]{hm}
Haardt, F., \& Madau, P. 1996, \apj, 461, 20

\bibitem[Kennefick, Djorgovski, \& {de  Carvalho} (1995)]{kennefick95b}
Kennefick, J.~D., Djorgovski, S.~G., \& {de Carvalho}, R.~R. 1995, \apj, 110,  
2553

\bibitem[Massey \& Gronwall (1990)]{massey90}
Massey, P., \& Gronwall, C. 1990, \apj, 358, 344

\bibitem[{Meiksin} (1994)]{meiksin94}
{Meiksin}, A. 1994, \apj, 431, 109

\bibitem[Oke (1974)]{oke74}
Oke, J.~B. 1974, \apjs, 27, 21

\bibitem[Oke (1990)]{oke90}
---------. 1990, \aj, 99, 1621

\bibitem[Oke, Cohen, Carr, Cromer, Dingizian, Harris,  Labrecque, Lucinio,
Schaal, Epps, \& Miller (1995)]{lris} 
Oke, J.~B.,  et~al. 1995, \pasp, 107, 375

\bibitem[Oke \& Korycanski (1982)]{okekor}
Oke, J.~B., \& Korycanski, D.~G. 1982, \apj, 255, 11

\bibitem[Reisenegger \& {Miralda-Escud\'e} (1995)]{reisen95}
Reisenegger, A., \& {Miralda-Escud\'e}, J. 1995, \apj, 449, 476

\bibitem[Scheuer (1965)]{scheuer65}
Scheuer, P. A.~G. 1965, \nat, 207, 963

\bibitem[Schlegel, Finkbeiner, \& Davis (1998)]{schlegel98}
Schlegel, D.~J., Finkbeiner, D.~P., \& Davis, M. 1998, \apj, 500, 525

\bibitem[Schneider, Schmidt, \&  Gunn (1991a)]{ssg91a}
Schneider, D.~P., Schmidt, M., \& Gunn, J.~E. 1991a, \aj, 101,  2004

\bibitem[Schneider, Schmidt, \&  Gunn (1991b)]{ssg91b}
---------. 1991b, \aj, 102, 837

\bibitem[Shklovskij (1964)]{shklov64}
Shklovskij, I.~S. 1964, Astron. Tsirkulyar, AN SSR, 3, 303

\bibitem[Songaila, Wampler, \& Cowie (1997)]{deut_limits}
Songaila, A., Wampler, E.~J., \& Cowie, L.~L. 1997, \nat, 385, 137,  
astro-ph/9611143

\bibitem[Steidel, Dickinson, \& Persson (1994)]{stei94}
Steidel, C.~C., Dickinson, M., \& Persson, S.~E. 1994, \apj, 437, L75

\bibitem[Stone (1977)]{stone77}
Stone, R. P.~S. 1977, \apj, 218, 767

\bibitem[Vogt, Allen, Bigelow, Bresee, Brown, Cantrall,  Conrad, Couture,
Delaney, Epps, \& Hilyard (1994)]{vogt} 
Vogt, S.~S.,  et~al. 1994, S.P.I.E., 2198, 362

\bibitem[Webb, Carswell, Lanzetta, Ferlet, Lemoine, Vidal-Madjar \& Bowen
(1997)]{webb} 
Webb, J.~K. et~al. 1997,  Nature 388, 250

\bibitem[Williger, Baldino, Carswell, {Cooke}, Hazard,  Irwin, McMahon, \&
{Storrie-Lombardi} (1994)]{williger94} 
Williger, G.~M., Baldino, J.~A., Carswell, R.~F., {Cooke}, A.~J., Hazard, C.,
Irwin, M.~J., McMahon, R.~G., \& {Storrie-Lombardi}, L.~J. 1994, \apj, 428, 574

\bibitem[Zhang, Anninos, \& Norman (1995)]{zhang95}
Zhang, Y., Anninos, P., \& Norman, M.~L. 1995, \apjl, 453, L57

\bibitem[Zhang, Anninos, Norman, \& Meiksen (1997)]{zhang97}
Zhang, Y., Anninos, P., Norman, M.~L., \& Meiksen, A. 1997, \apj, 485, 496

\bibitem[Zhang, Meiksen, Anninos, \& Norman (1998)]{zhang98}
Zhang, Y., Meiksen, A., Anninos, P., \& Norman, M.~L. 1998, \apj, 495, 63

\end{thebibliography}
\end{document}